\documentclass[aps,preprint,noshowpacs,prb,preprintnumbers]{revtex4-1}
\usepackage[latin1]{inputenc}
\usepackage{dcolumn}
\usepackage{graphicx}

\begin{document}

\title{ Absence of superheating for ice $I_{h}$ with a free surface : a new method 
of determining the melting point of different water models. }

\author{ Carlos Vega and  Maria Martin-Conde }
\affiliation{Departamento de Quimica Fisica, Facultad de Ciencias 
Quimicas, Universidad Complutense, 28040, Madrid, SPAIN }
\author{  Andrzej Patrykiejew   }
\affiliation{ Faculty of Chemistry, MCS University, 20031, Lublin, POLAND  }

\date{\today}
\begin{abstract}
Cite as : C. Vega, M.Martin-Conde and A.Patrykiejew, Mol.Phys., 104,  3583,  (2006). \\
Molecular dynamic simulations were performed for 
ice $I_{h}$ with a free surface. 
The simulations were carried out at several temperatures and each run lasted 
more than 7ns. At high temperatures
the ice melts. It is demonstrated that the melting process starts at the  
surface and propagates to the bulk of the ice block. Already at the temperatures 
below the melting point, we observe a thin liquid 
layer at the ice surface, but the block of ice remains stable along
the run. As soon as the temperature reaches the melting point the entire ice block
melts. Our results demonstrate that, unlike in the case of conventional simulations
in the NpT ensemble, overheating of the ice $I_{h}$ with a free surface does not
occur. That allows to estimate the melting point of ice at zero pressure.
We applied the method to the 
following models of water: SPC/E, TIP4P, TIP4P/Ew, TIP4P/Ice and
TIP4P/2005, and found  good agreement between the melting temperatures obtained by 
this procedure and the values obtained either from free energy calculations or 
from direct simulations of the ice/water interface. 
\end{abstract}
\maketitle

\section{INTRODUCTION}
  For obvious reasons, liquid water has been the focus of thousands
of simulation studies since the pioneering work of Barker 
and Watts\cite{barker69} 
and  Rahman and Stillinger\cite{JCP_1971_55_03336}. However, simulation studies of the 
solid phases of water (ices) have
been much more scarce\cite{JCP_1982_76_00650,baez95,ayala03,baranyai05,rick05,rusosreview}. There 
is a number of reasons to study solid
and amorphous phases of water by computer simulation methods. 
First of all, the description of the phase diagram represents
a major challenge for any potential model of water. Secondly, many 
experimental facts concerning ices and amorphous water are not 
completely understood and computer simulations could help in 
obtaining a molecular view of the process. Just to show a 
few examples, let us mention the nucleation of ice \cite{trout03,matsumoto02},
the dynamics of solid-solid
transitions, the possibility of liquid-liquid equilibria
\cite{poole92,mishima98,debenedetti03,brovchenko05},
the speed and mechanism of crystal growth \cite{nadagrowth,carignano05}
and  the properties of 
ice at the free surface\cite{kroes92,nada97,nada00,kawamura04,carignano05}. 

Any researcher performing simulations of water
must choose a potential model among the many now 
available\cite{JML_2002_101_0219,chaplinweb,jorgensen05}.  
The most popular were developed in 
the eighties and are known as: TIP3P\cite{jorgensen83} , TIP4P\cite{jorgensen83,bernal33}, 
TIP5P\cite{mahoney00}, SPC\cite{berendsen82} and SPC/E \cite{berendsen87}. 
They were fitted to reproduce the properties of liquid water at room temperature and
pressure. Before choosing a potential model of water it seems quite reasonable to 
ask about its ability to describe the phase diagram of water. 
Over the last decade, the vapour -- liquid coexistence has been determined
for several water models by using the Gibbs 
ensemble technique\cite{boulougouris98,panagiotopoulos98,lisal01,lisal02,siepmann2000,siepmann05,saintmartin2005,vega06}.
The critical properties of many water models  are now known.
However, the melting point of these models has been studied less
often. Haymet et al. provided an estimate of the melting point of SPC, TIP4P and 
SPC/E water models\cite{karim88,karim90,bryk02}, by performing simulations of the liquid-solid interface.
The same approach has been followed recently by Wang et al.\cite{wang05} and by Nada and Furukawa\cite{nada05}. 
By performing free energy calculations the melting points of
the models SPC/E, TIP4P ,TIP5P and NvdE\cite{JCP_2003_118_07401} have been 
reported\cite{baez95,gao00,tanaka04,JCP_1999_110_00055}.
It should be stated that estimates of the melting point of ice
obtained by different authors and different techniques (or even by the same authors using 
different techniques) do not always agree, so it is still of interest
to determine the melting point by as many different routes
and methodologies as possible. 
Over the last few years, we have undertaken a systematic study of the solid 
phases of water\cite{sanz1,sanz2,vegafilosofico,pccpgdr}.
Firstly the phase diagrams for SPC/E and TIP4P models were obtained from
free energy calculations \cite{sanz1,sanz2} by using the 
methodology proposed by Frenkel and Ladd\cite{frenkel84}, and extended to water 
models by Vega and Monson \cite{vega98}. Secondly, by using the Hamiltonian Gibbs-Duhem integration\cite{singer90,kofke93,ACP_1998_105_0405_nolotengo,ACP_2000_115_0113_nolotengo,laird}  
 the melting point of ice $I_{h}$ was obtained for several 
models of water \cite{vegafilosofico}. In these two last cases we used Monte Carlo methods
and "home-made" codes. Then, we applied a completely independent methodology, namely molecular dynamics method\cite{ramon06}, and used the GROMACS package\cite{gromacs33} to
perform  direct simulations of  fluid-solid coexistence\cite{ladd77,ladd78}.
It should be emphasized that we obtained excellent agreement between  
the normal melting temperatures $T_{m}$ ( i.e the melting temperature at the normal pressure of $p=1bar$ ) estimated by different routes. 
That was the case for the traditional SPC/E, TIP4P and TIP5P, and also for the new
generation of models proposed just in the last two years, namely TIP4P/Ew\cite{horn04}, 
TIP4P/Ice\cite{tip4pice} and 
TIP4P/2005\cite{abascal05}, designed to improve the description of ices and water. 
Of course, each model predicts a different normal melting temperature, and the 
TIP4P/Ice\cite{tip4pice} model is known to perform best leading to $T_{m} = 270\pm 2$K.

One may  wonder why it is so cumbersome to determine the melting point
of a water model. After all, melting points are determined easily 
in the lab conditions for many substances by simply heating the solid at constant
pressure until it melts. 
Although liquids can  be supercooled, solid can not be superheated 
as first stated by Bridgmann \cite{PAAAS_1912_XLVII_13_441_photocopy}:
" It is impossible to superheat a crystalline phase with respect to 
the liquid. ".
For this reason there is in principle no risk in determining 
melting points by simply heating the solid until it melts. 
However, in computer simulations the situation is somewhat 
different. If conventional NpT simulations are performed for ice $I_{h}$ then 
it does not melt at $T_{m}$ but at a somewhat higher 
temperature \cite{mcbride05,mcbride04,gay02,luo04}.
It is worth mentioning that superheating has been also recently found 
 experimentally\cite{superheating_1,superheating_2} 
for ice $I_{h}$ but only on a short 
time scale (about 200ps). 
The difference between the results of NpT simulations and 
those found in experiments (summarized in the Bridgman's 
statement) is striking. However, there is  a fundamental difference 
between experiments and NpT simulations of bulk solids. Whereas in 
experiments the ice (or a solid in general) must have a surface (or rather
an interface at which it is in a contact with another phase; vapour or liquid), 
this interface is missing in conventional NpT simulations of bulk solids.
When no interface is present ice must melt via bulk melting\cite{parrinellobulkmelting}.

It is now commonly accepted that melting starts at the surface and already at the
temperatures lower than the bulk melting point, solids exhibit a liquid-like layer
at the surface\cite{abraham1981,van-der-Veen-at.all_1988,Dash_89,Bienfait_92,broughton1983}. 
Different melting scenarios are possible depending on the behaviour of that liquid-like
layer when the bulk melting temperature is approached from below. When the thickness
of the liquid-like layer increases and finely diverges at $T=T_{m}$, i.e., when the
solid is wetted by its own melt, one meets the so-called surface induced melting or just surface melting \cite{van-der-Veen-at.all_1988,Dash_89,Fr_Pin_King_93,Nenov_84,Fran_Veen_85}.
On the other hand when the liquid-like layer retains finite thickness up to $T=T_{m}$, i.e., when the
melt does not wet completely the solid surface (partial wetting), one meets the case of surface nonmelting\cite{Carnaveli_etal_87,Tartaglino_etal_05}.  
Both situations have been observed experimentally and found in computer simulations (for a 
review see that of Tartaglino et al.\cite{Tartaglino_etal_05}). No matter which of the two above mentioned mechanisms occurs,
the existence of an interface prevents the solid from overheating.
One may therefore expect that the presence of
a free surface at the ice sample, would turn the results of computer simulations 
back to normal, i.e. with melting occurring right at the bulk melting temperature. 

  The purpose of this paper is two-fold. On one hand we would like to check whether
a piece of ice with a free surface can or not be superheated in computer
simulation. 
As it will be shown shortly, the answer to this question is that 
when a free surface is present, superheating is suppressed in computer
simulation, just the same as in experiment.  Things go back
to normal, and Bridgman's statement holds again. 
Secondly the suppression of superheating means that there is another 
relatively straightforward way to determine the normal melting 
point.  As it will be shown , the melting point obtained from the simulation 
of the free surface agrees with the estimates obtained by other routes.
The problem of estimating melting points of water models seems solved,
remaining uncertainties being of the order of about 3-4K.

\section{METHODOLOGY}
Let us first describe the procedure used to get the initial
configuration.  Although the ice $I_{h}$ 
is hexagonal, it is possible to use an orthorrombic unit cell\cite{petrenko99}.
It was with this orthorrombic unit cell that we generated the initial
slab of ice.  In ice $I_{h}$,  protons are 
disordered whereas still fulfilling the Bernal-Fowler rules 
\cite{bernal33,pauling35}. We used the algorithm of 
Buch et al.\cite{buch98} to obtain 
an initial configuration with proton disorder and  almost zero dipole
moment (less than 0.1 Debye). This initial configuration contained 
1024 molecules of water and fulfilled the Bernal Fowler rules.
 The dimensions of the ice sample were 
$36$ ~\AA\    x $31$ ~\AA\ x $29$ ~\AA\ . 
In oreder to equilibrate the solid, NpT simulations of bulk ice $I_{h}$ were performed at zero pressure 
at each temperature of interest.  
We used the molecular dynamics package Gromacs (version 3.3)\cite{gromacs33}.
The time step was 1fs and the geometry
of the water molecules was enforced using constraints\cite{shake,berendsen84}.
The Lennard-Jones of the potential (LJ) was truncated at 9.0~\AA\ . 
Ewald sums were used to deal with electrostatics.
The real part of the coulombic potential was truncated at 9.0~\AA\ .
The Fourier part of the Ewald sums was evaluated by using the Particle Mesh
Ewald (PME) method of Essmann {\em et al.}\cite{essmann95}.
The width of the mesh was 1~\AA\ and we used a fourth other
polynomial.  The temperature was kept by using a 
Nose-Hoover\cite{nose84,hoover85} thermostat with a relaxation time of 2~ps.
To keep the pressure constant, a Parrinello-Rahman 
barostat\cite{parrinello81,nose83} was used.  The relaxation time of the barostat was of 2~ps.
The pressure of the barostat was set to zero. 
We used  a Parrinello-Rahman barostat where all three sides 
of the simulation box were allowed
to fluctuate independently. The angles were kept orthogonal during this NpT 
run, so that they were not modified with respect to the initial configuration. 
The use of a barostast allowing independent
fluctuations of the lengths of the simulation box is important.
In this way, the solid  can attain the
equilibrium unit cell for the considered model and thermodynamic conditions. It is not a good idea to 
impose the geometry of the unit cell. The system should rather determine it from NpT runs. 
Once the ice is equilibrated at zero pressure we proceed to generate the ice-vacuum interface. 
By convention, we shall assume in this paper that the x axis is perpendicular to the
ice-vacuum interface. The ice-vacuum configuration was prepared by simply changing the 
box dimension in the x axis from its value ( around 36 ~\AA\ ) to 100 ~\AA\  .
As a result, the dimensions
of the simulation box are now 100 ~\AA\ x 31 ~\AA\ x  29 ~\AA\ (the y and z are approximate since the actual 
values were obtained from the NpT runs for each model and at any particular value of the
temperature).  The approximate area of the ice-vacuum 
interface was 31 ~\AA\ x 29 ~\AA\ (approximately).  This is about 10 molecular diameters in each 
direction parallel to the interface. In this work we have chosen the x axis to be of 
the 1\={2}10  direction.
In other words, the ice is exposing its  prismatic secondary plane 
to the vacuum  (i.e the 1\={2}10 plane). 
In fig.1 we show the initial configuration of a perfect ice as seen from the basal plane. 
More details regarding ther relation of the main planes of ice (basal, primary prismatic and secondary prismatic) 
to the hexagonal unit cell can be found in 
in figure 1 of the paper by Nada and Furukawa \cite{nada05}, in the paper by 
Carignano et al.\cite{carignano05} and also in the water web site of Chaplin \cite{chaplin}.
Obviously, the properties of the ice-vacuum interface depend on the crystal 
plane facing the solid-vacuum 
interface. In particular, the melting mechanism (surface melting or surface nonmelting) may depend on the structure
of the surface plane.
On the other hand, the melting temperature should not depend on the choice of the plane 
selected to form the ice-vacuum interface. Therefore, the choice of the prismatic secondary
plane is a good as any other. 
The main advantage arises from the fact that for the ice-water interfaces it has been shown 
that the 1\={2}10 plane exhibits the fastest dynamics \cite{nada05,carignano05}.
It is also a convenient face to make a visualization of the melting
process.
Once the ice-vacuum system was prepared, we performed relatively long NVT runs (the lengths was 
between 4ns and 14ns  depending on the model and thermodynamic conditions). Since 
we have been using NVT molecular dynamics, the dimensions of the simulation box 
have been fixed, of course, unlike in the preceding NpT run.

\section{RESULTS}
 Let us first present results for the TIP4P model. In fig.2 the results 
of the total energy  along the  run are presented. 
As it is seen the total energy presents two different behaviours depending on 
the temperature. At high temperatures, the total energy of the system increases
continuously and then reaches a plateau. The visualization of particle trajectories indicates
that the block of ice melts and finally one obtains an slab of liquid in equilibrium
with its vapour. The melting speed depends on the temperature but as it can be seen, 
close to the model melting point it is possible to melt the ice completely in about 6ns. 
The 
difference between the initial and the final energy is about 3.5 kJ/mol, indicating 
that the melting enthalpy is roughly of this order of magnitude. This is in reasonable
agreement with the melting enthalpy of the TIP4P model, estimated in our previous
work\cite{vegafilosofico} for the melting of bulk ice to bulk water, which was of 
about 4kJ/mol (the somewhat smaller value found here is due to the presence of the free 
surface ). Since the block of ice
is of a thickness of about 36 ~\AA\ , one can approximately state that an increase of energy by
about 1 kJ/mol corresponds to a decrease of the thickness of the ice block by about 10 ~\AA\ .
Taking into account that we have two solid/vacuum interfaces, one at the right
and the other at the left side of the ice block, it means roughly speaking an increase
of the total energy by 1kJ/mol, which amounts to the formation of a liquid layer of about 
5~\AA\ (the argument is not elaborated but at least provides some orders of magnitude).
The behaviour at low temperatures is different. At the beginning (first 1-2ns), 
there is an increase of the
energy but after that the energy remains approximately constant, apart from the thermal fluctuations.
The analysis of the 
configurations of the TIP4P at $T=228K$, shows that the increase of energy during the first 1ns 
is due to the formation of a thin liquid layer at the surface of
ice, which may indicate the onset of surface melting, mentioned already in the Introduction,
and first proposed by Faraday \cite{physics_today}. 

We just recall that the term 
"surface melting" usually represents the formation of 
a quasi-liquid layer (qll) on the surface of a solid at temperatures still below the 
melting point. The properties of this quasi-liquid layer are similar to, but not identical with, those
of a bulk fluid under the same conditions. The thickness of the 
layer depends on the particular plane forming the solid-fluid interface 
and on the proximity to the melting point. It may diverge to infinity or stay finite as the temperature
approaches $T_{m}$ from below.  The surface melting of ice, has been found both 
in experiment (see \cite{dash,henson} and references therein ) and
in computer computer simulation for several potential models of 
water\cite{kroes92,nada97,nada00,kawamura04,carignano05}.
Some theories are also available to explain its origin\cite{dash,henson}.
One should note that the surface melting is an equilibrium phenomenon, leading to the lowering of 
the system free energy by replacing the ice-vacuum interface by the ice-liquid* and liquid*-vapour interfaces.
By liquid* we mean a quasi-liquid layer of microscopic
thickness. The existence of a quasi-liquid layer at temperatures below the
melting point is not exclusive of ice, but it does also appear in simpler
liquids such as metals\cite{Fran_Veen_85,pluis1987} or Lennard-Jones 
models\cite{abraham1981,broughton1983}. The study of the quasi-liquid layer thickness by computer simulation 
is of interest by its own.
This is so because different experimental techniques provide quite different values of its  
thickness\cite{dash,henson} and it is of interest to understand why this is the case. 
We expect to study that  in future work. For the purpose of this work
it is just enough to state, that at low temperatures a qll appears at the ice surface, provoking 
an increase of energy in the first 1-2ns of the run, but then the energy remains constant (with 
the normal thermal fluctuations) after this initial period. In that respect no signal of melting 
of the block of ice has been observed at the temperatures below $T_{m}$.
By repeating the simulation at several temperatures it is possible to determine the lowest 
temperature at which the block of ice melts $T_{+}$, and the highest temperature at which it does not 
melt $T_{-}$. By taking the average of these two temperatures we obtain what we call  
$T_{s}= 1/2 ( T_{+} + T_{-} ) $. 
For the length of the runs used (about 4-10ns) the typical difference between $T_{+}$ and
$T_{-}$ is about 3-4K. It would be possible to reduce this temperature window by performing 
longer runs (of the order of hundred ns rather than of 10ns). However, taking into account that
the current uncertainty in the estimations of the water melting point for different models 
obtained from the free energy calculations is just about 4K, the accuracy obtained by the runs
 of the length presented here seems to be quite 
sufficient. The crucial question now is, what $T_{s}$ actually is?
 How does $T_{s}$ compare with the melting point $T_{m}$?
For the case of TIP4P, the melting point has been determined by several  groups. From the free energy 
calculations we obtained\cite{sanz1}, $T_{m}=232K$. Starting from the melting 
point of the SPC/E model obtained by free energy calculations ( $T_{m}=215K$ ) and 
using  Hamiltonian Gibbs-Duhem integration we obtained again $T_{m}=232K$ for 
the TIP4P model\cite{vegafilosofico}. 
Using  free energy calculations, Koyama et al.\cite{tanaka04} 
have obtained $T_{m}=229K$. From direct coexistence between the fluid and the solid phase
we have obtained $T_{m}=228K$ \cite{ramon06}. From direct coexistence 
between the fluid and the solid 
Wang et al.\cite{wang05} have obtained $T_{m}=229K$. From the results presented in 
four different papers, each using different
techniques and methods, it seems clear that the melting point of the TIP4P model is about 
$T=230(3)K$. This is also consistent with a corresponding states rule found by us recently. 
We have found\cite{vega06} that for TIP4P model the ratio of 
the melting point to the critical point temperatures
is about 0.394. This estimate of the melting point temperature is consistent with the 
corresponding states rule and with the well known value of the critical temperature of 
the TIP4P model\cite{lisal01,baranyai06,vega06}. 
The fact that there are some small discrepancies between different estimates is not surprising.
They are due to the fact that long range forces are treated in a different way in different 
papers (truncation of the potential versus Ewald sums), different ice configurations (recall the 
issue of the proton disorder), different cutoff of the potential, as well as possible finite-size 
effects. Therefore, it can be stated with confidence that 
the melting point of the TIP4P is about $T=230(3)K$.
The value of $T_{s}$ found in this work is $T_{s}=229(2)K$. The error in
$T_{s}$ is mainly determined by the grid of temperature used, 
being the difference between $T_{+}$ and 
$T_{-}$ of about 4K, so that the error in $T_{s}$ is of about 2K. 
The obvious conclusion is that for the TIP4P model, and within the uncertainty 
of different estimates, $T_{s}$ is identical to 
$T_{m}$. In other words, overheating has been completely suppressed by the presence of a free surface.

 In order to test the hypothesis, stating that $T_{s}$ and $T_{m}$ are identical, we 
have performed
runs for other different models of water, namely the recently proposed 
TIP4P/Ice\cite{tip4pice} , TIP4P/2005\cite{abascal05} , TIP4P/Ew\cite{horn04} and the classical  SPC/E model.  Results of 
the runs for these models are presented in 
Figs.3-6 respectively. In Table I, the values of $T_{s}$ for different 
models are presented and compared with the values of $T_{m}$ obtained from 
a direct study of a fluid-solid coexistence,
as well as from  the Hamiltonian Gibbs-Duhem integration. As it can be seen, 
$T_{s}$ is identical to $T_{m}$, within the error bar. This result seems to be important, since it shows 
that the similarity
between $T_{s}$ and $T_{m}$ is not a particular feature of the TIP4P model, but applies to
any other model of water. The conclusion is that when a free surface 
is available overheating of the solid is 
suppressed, and the ice melts at the melting temperature, provided the length of the 
runs are sufficient (of the order of 10ns or larger), and the size of the ice block is sufficiently large,
not smaller than about 3-4nm. 
To check whether this conclusion depends on the system size and/or of the
face exposed to vacuum we have performed a run for the TIP4P/Ice model at 
a temperature of T=274K (higher than the melting point). In this particular case we used
1536 molecules and the plane at the ice-vacuum interface was the primary
prismatic plane. The area of the interface was of about 
$27$ ~\AA\    x $30$ ~\AA\  . The width of the ice was  of $63 $ ~\AA\ approximately. 
Again, complete melting of the sample was observed, indicating 
that the physics was not changed neither  by a larger system nor by a different
plane exposed. However, in this case it took about 25ns to melt the ice completely
due to a larger size of the system and to slower dynamics.

A minor comment is in order here. The melting point obtained in our
previous work $T_{m}$ corresponds to the melting point at the normal pressure
$p=1 bar$. However, $T_{s}$ was obtained in this work from simulations at zero pressure.
Obviously, the melting temperature at zero pressure and at the pressure of 
one bar differ somewhat. In fact, for real water the difference is  
about 0.01K (i.e 273.16K for essentially zero pressure at the triple point versus
273.15K for $p=1 bar$). For water models, the difference between the melting 
pressure at zero pressure and at $p=1 bar$ is expected to be of the order 
of $1/(dp/dT)$. By using the values of $dp/dT$ reported in our 
previous work\cite{vegafilosofico} it can be seen that also for water models 
the difference between the melting
temperatures at normal pressure and zero pressure should be of the order of 
about 0.01K (i.e quite small). In view of that, it it reasonable to identify
$T_{m}$ with $T_{s}$. Still another comment is required with respect to 
our simulations of the ice-vacuum interface. 
In principle, when ice is exposed to vacuum at temperatures below the melting
point there should be a solid-vapour equilibria. The vapour pressure of water
at the triple point is quite low, namely $6\times 10^{-3}$ bar. The vapor
pressure of TIP4P models at the triple point is fifty times smaller\cite{vega06}, 
about $1.3\times 10^{-4}$ bar. Although in principle one
could determine the vapour pressure of ice by performing NVT runs of ice
in contact with vacuum, this is not feasible in practice. In fact, for 
the lengths of the runs used in this paper we have never observed the 
sublimation of a single  ice molecule into the vacuum (this of course should
occur in very long runs, but since it is a rare event it was not observed 
in the runs of this work which lasted about 10ns). For this reason 
we prefer to say that $T_{s}$ was obtained at zero pressure, but 
it could be more appropriate
to say that it was obtained at the sublimation pressure of the ice (which
must be quite low anyway). 

  It is interesting to see the mechanism of the melting process.
We shall present results for the TIP4P/Ice at $T=276K$ (a temperature
higher than the melting point of the model). In fig.7, a snapshot of the system after
1ns is shown. In fig.8, the final snapshot of the system after 5ns of
run is shown. As shown in fig.8, the ice has melted completely after the 
run of 5ns. 
From the results given in Fig.7 it is also clear that the melting process starts at the 
surface and then  propagates to the interior or the block of ice. We have analyzed 
the movies of all the runs and always found this to be the case. We have never 
observed the formation of a droplet of water in the
interior part of the ice. 
We have always observed that the melting 
started at the surface and propagated to the interior part of the block of ice. 
One of the first methodologies proposed to estimate the melting point 
in computer simulations was the direct 
coexistence method\cite{ladd77,ladd78,cape78}. In this method,
the solid is put into contact with the liquid at a certain temperature and pressure. 
If the temperature is higher than the melting 
temperature, the solid part of the sample will melt. If the temperature is lower than 
the melting temperature then the liquid will freeze.
Regarding the ice-water interface, it has been investigated during the last
fifteen years and estimates of the melting points for several water models have
been given.\cite{karim88,karim90,bryk02,bryk04,wang05,nada05,carignano05,ramon06}.
Recently, we have found that melting temperatures estimated from the direct 
coexistence method are in agreement with those obtained from free energy calculations.
The presence of a "nucleus" of ice in the sample, and of a "nucleus" of liquid in 
the sample prevents the possibility of metastability. In other words, the presence 
of the two phases in the simulation sample from the very beginning avoids the 
possibility of metastability. The metastability is due to the fact that the formation 
of a nucleus of a new phase in the interior of another phase is an activated process. 
Although this is pretty clear, it was not so obvious what happened  when the free surface
of the solid was heated while exposed to vacuum. This work shows that also in this 
case superheating is suppressed. To prove further that a liquid layer is
already present in the sample at temperatures well below the melting point,
we present in fig.9 the final configuration (after a 8ns run)
 obtained for the TIP4P/Ice at 
a temperature well below the melting point of the model ($T=264K$). As it
can be seen, a quasi liquid layer is already present in the system.
In fig.10 the final configuration (after a 9ns run) obtained for the TIP4P/Ice
at a temperature just below the melting temperature $T=270K$ is shown.
It can be observed that the thickness of the quasi-liquid layer is larger now 
that at the lower temperature. The question whether the thickness of the liquid 
layer diverges when the melting point  is approached (total wetting) or 
remains finite (partial wetting) should be studied in more detail.
 In any case figures 9 and 10 illustrate clearly the point that 
a quasi-liquid layer is already present in ice at temperatures below the
melting point so that superheating is suppressed.

\section{Conclusions}
 In this work, we have performed NVT Molecular Dynamics simulations of ice $I_{h}$ with a free surface. The ice was first equilibrated by performing NpT
simulations at  zero pressure. Simulations were then performed at several 
temperatures and runs lasted between 5 and 10ns. 
At  temperatures below the melting point,
the ice surface develops a thin liquid like layer (quasi liquid 
layer) of microscopic thickness. 
However, at temperatures above the melting point the ice melt, and does not exhibit any trace
of overheating.  In all cases, we have observed that the melting process 
starts at the surface and propagates to the interior of the material. 
We have estimated the melting point of ice $I_{h}$  
for several water models, namely ,
TIP4P, TIP4P/Ice, TIP4P/2005, TIP4P/Ew, and SPC/E. 
Since the melting points of these water models had been evaluated previously 
(for ice $I_{h}$)
by free energy calculations and by fluid-solid interface simulations, it is of interest
to compare to the values obtained in this work. Excellent agreement has been found
(see Table I) between previous estimates of $T_{m}$ and those obtained in this 
work from simulations of the free  surface.  
Therefore, although it is possible to
overheat ice in the usual BULK NpT simulation\cite{mcbride05,mcbride04}, the 
overheating does not seem to be possible as soon as the ice  has a free surface.
Although this phenomenon has been studied before 
for spherical fluids\cite{Tartaglino_etal_05,Boutin_93}, to the best
of our knowledge this is the first time that the issue is addressed for
such a complex fluid as water. 
As a bonus, one has a remarkably simple method to estimate the melting 
temperature of a water model: equilibrate first the material in 
an NpT run, and perform afterwards a long  NVT run (of about 10ns or
more). This is a long run but still within the capacity of current
computers.  Taking into account that estimating melting temperatures by 
free energy calculations is somewhat involved (although feasible) the 
existence of an alternative and simpler method is welcome. One should note that
the method is probably not limited to water, but can also be used for 
other complex molecules. The method can be applied only when the following 
two conditions are met. Firstly, the melting of the system at temperatures above
the melting point should be fast enough to be studied within a reasonable time by 
molecular simulations. Secondly, the liquid layer thickness at the temperatures below,
but not too far from the melting point, should not be too large. Otherwise, the 
finite size of the sample may lead to the observation of complete melting already
below the melting point. 
It should be pointed out that the procedure allows to estimate the zero pressure melting 
temperature, which should be very close to the normal melting point, but does not
allow to estimate the fluid-solid coexistence at higher pressures.  

  Now, back to the classroom you may explain that the fluid-solid transition
(in the absence of impurities leading to heterogeneous nucleation) occurs
by the following mechanism: nucleation (of an embryo of the solid phase) and 
growth.  Liquid water can be supercooled  because the formation of an embryo is 
an activated process and that requires a certain amount of time to occur.
You may now explain that melting occurs by the same mechanism: nucleation
(of am embryo of the liquid phase) and growth. The point is that for ice
with a free surface the embryo of the liquid phase 
is already there, at the surface of the solid. For this reason, 
the activation energy of forming the liquid embryo is  zero
and melting is just  the growth of a new
phase (i.e the propagation of the liquid layer from the surface
to the interior of the solid). 
 As stated by Bridgmann in his classical 1912 
paper \cite{PAAAS_1912_XLVII_13_441_photocopy}: 
" It is impossible to superheat a crystalline phase with respect to the liquid.
No good reason for this has ever been given, but no exception has ever
been found, and it is coming to be regarded as a law of nature ".
We agree except for the absence of explanation for this behaviour.
As stated by Frenkel\cite{frenkel_kinetics,dash} and suggested by 
Tammann\cite{tammann1910}: " It is well known
that under ordinary conditions an overheating of a crystal, similar
to the overheating of a liquid, is impossible. This peculiarity is
connected with the fact that the melting of a crystal , which is 
kept at a homogeneous temperature, always begins on its free surface.
The role of the latter must, accordingly, consist in lowering the 
activation energy, which is necessary for the formation of a flat embryo
of the liquid phase, i.e. of a thin liquid film down to zero."

\section{acknowledgements}
It is a pleasure to acknowledge helpful discussions with 
Prof.J.L.F.Abascal, Dr.L.G.MacDowell, Dr.E.Sanz , Dr.C.McBride and 
R.G.Fernandez. 
This project has been financed by the grant FIS2004-06227-C02-02
of Direccion General de Investigacion,  by the  
project S-0505/ESP/0299 of the Comunidad de Madrid and by 
the European Community under the grant No. MTDK-CT-2004-509249. 
One of us (CV) would like to thank the group of Molecular Modeling
at the University of Lublin, and specially Andrzej Patrykiejew and Stefan Sokolowski 
for the hospitality during his stay in Lublin.

\newpage 
\begin{table}
\caption{Normal melting points $T_m$ of different water models as obtained from different 
         methodologies. The column labeled as free energy correspond to the 
         melting point either as obtained from free energy calculations\cite{sanz1,sanz2} 
         (SPC/E and TIP4P) or from Hamiltonian Gibbs Duhem integration\cite{vegafilosofico}.
         (TIP4P/Ice, TIP4P/2005, TIP4P/Ew).
         The results of the column labeled 
         as Solid-Liquid interface have been obtained from direct coexistence 
         of the solid with the liquid \cite{ramon06}.
         The last column is the value of $T_{s}$   
         (which is just the average of the lowest temperature at which ice melts and 
         the highest at which it does not melts) 
         obtained from simulations of ice $I_{h}$ with a free interface. The 
         similarity of $T_{s}$ and $T_m$ shows the absence of superheating for
         ice with a free surface. }
\begin{tabular}{ccccc}
\hline
Model& $T_{m}/K$ (Free energy) & $T_{m}/K$ (Solid-Liquid interface) & $T_{s}/K$ (this work)   \\
\hline
TIP4P       & 232(4)   & 229(2) & 230(2)    \\
TIP4P/Ice   & 272(6)   & 268(2) & 271(1)    \\
TIP4P/2005  & 252(6)   & 249(2) & 249(3)    \\
TIP4P/Ew    & 245.5(6) & 242(2) & 243(2)    \\
SPC/E       & 215(4)   & 213(2) & 217(2)   \\
\hline
\hline
\end{tabular}
\end{table}
\newpage 
\vspace*{2cm}
\begin{figure}[ht]
\caption{ Initial configuration of the ice $I_{h}$ with 1024 molecules. 
 The x axis goes from left to right so that
the ice-vacuum interface is located at the right and left sides. The 
plane that can be seen is the basal plane.  The plane exposed to the 
vacuum is the secondary prismatic plane.}  
\includegraphics[height=450pt,width=450pt,angle=0]{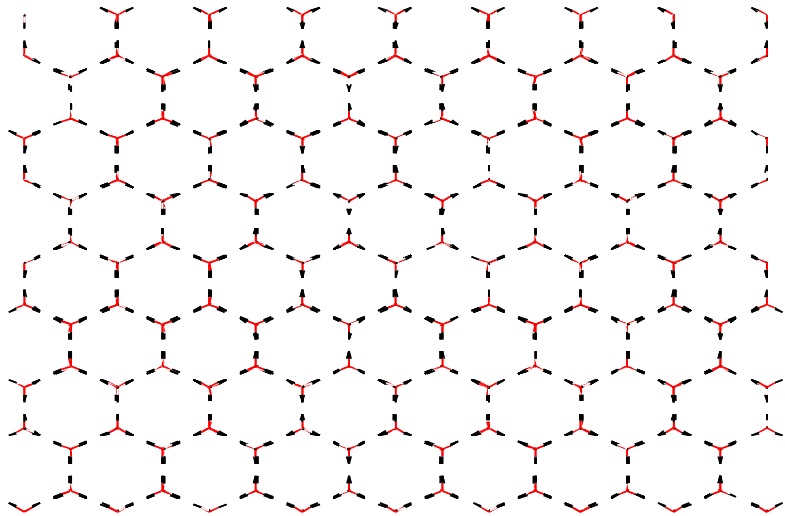}
\end{figure}

\newpage 
\vspace*{2cm}
\begin{figure}[ht]
\caption{ Total energy as a function of time obtained at several 
temperatures by performing MD simulations for a block of ice $I_h$ with a free 
surface. Results presented here correspond to the TIP4P model.}  
\includegraphics*[scale=0.45,angle=0]{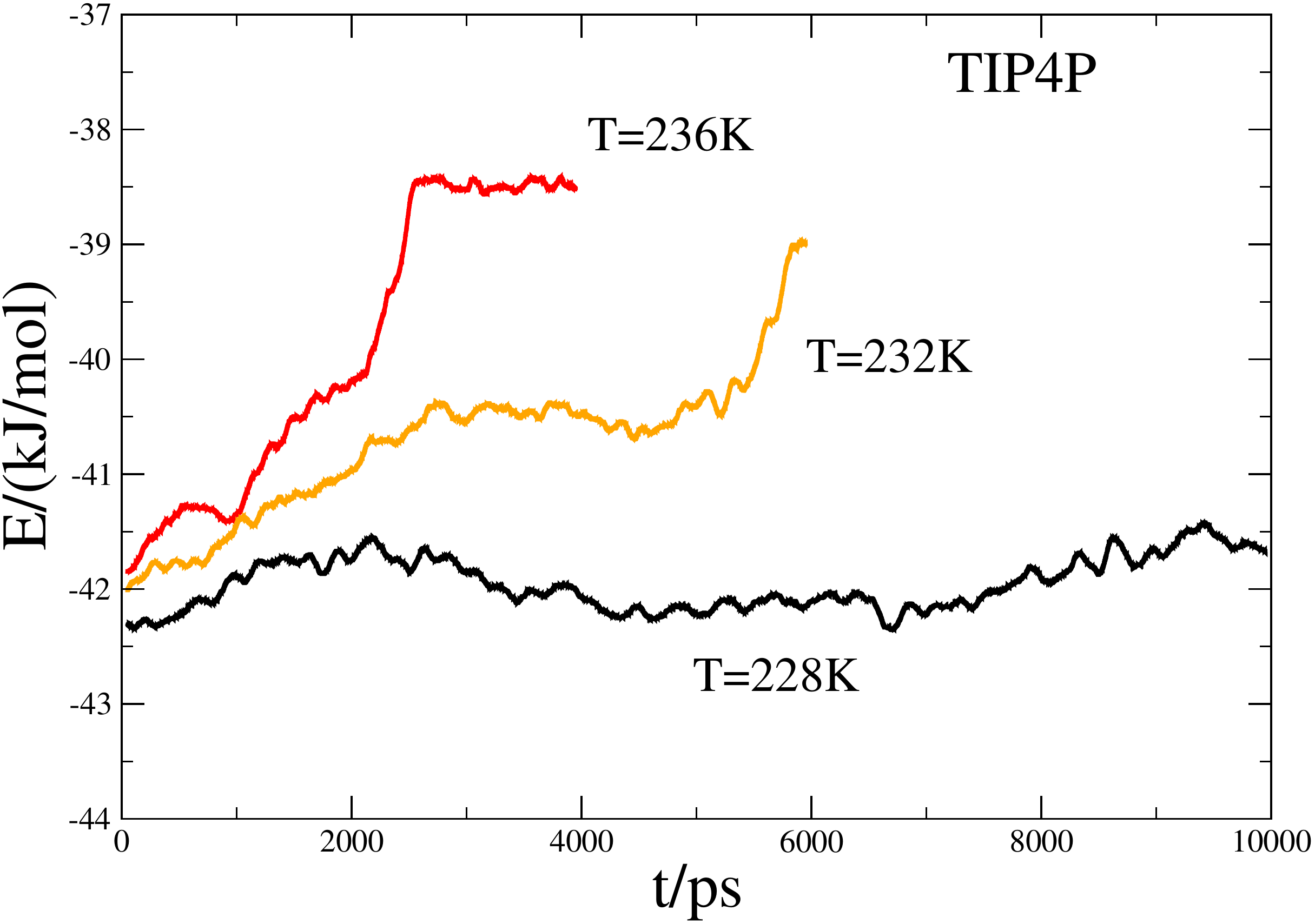} \\
\end{figure}

\newpage 
\begin{figure}[ht]
\caption{ Total energy as a function of time obtained at several 
temperatures by performing MD simulations for a block of ice $I_h$ with a free 
surface. Results presented here correspond to the $TIP4P/Ice$  model.}  
\includegraphics*[scale=0.45,angle=0]{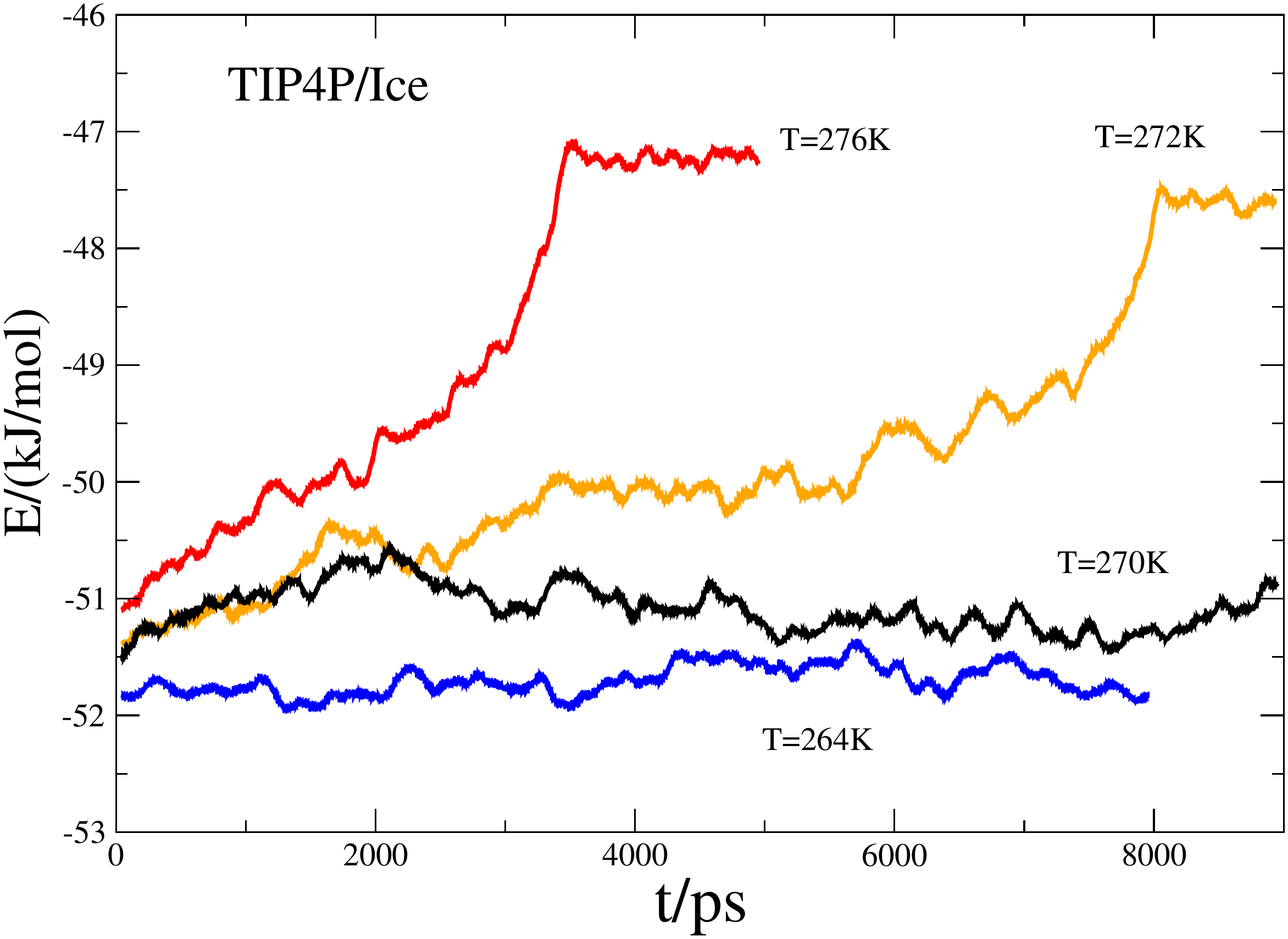} \\
\end{figure}

\newpage 
\begin{figure}[ht]
\caption{ Total energy as a function of time obtained at several 
temperatures by performing MD simulations for a block of ice $I_h$ with a free 
surface. Results presented here correspond to the TIP4P/2005  model.}  
\includegraphics*[scale=0.45,angle=0]{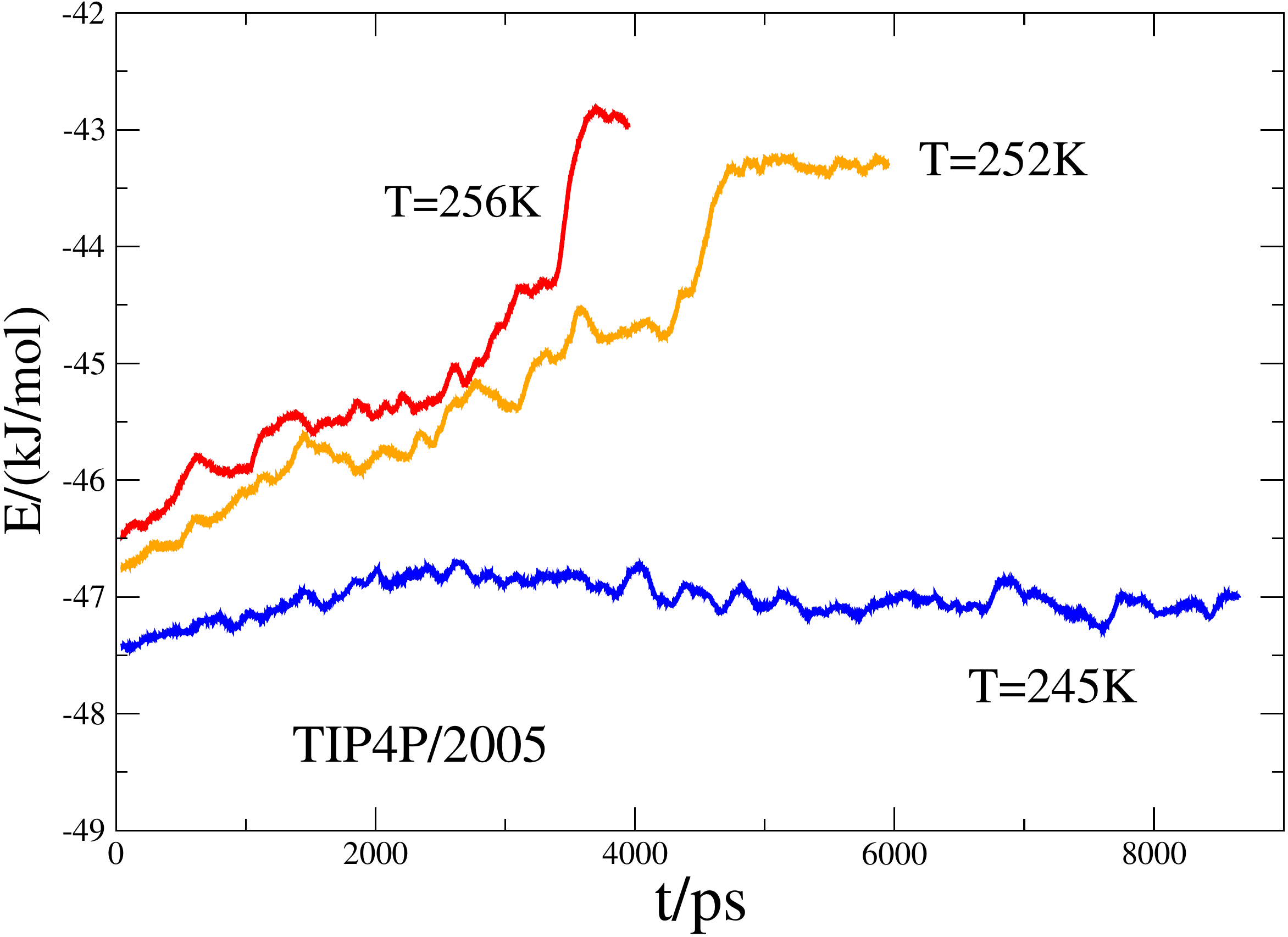} \\
\end{figure}
\newpage 
\begin{figure}[ht]
\caption{ Total energy as a function of time obtained at several 
temperatures by performing MD simulations for a block of ice $I_h$ with a free 
surface. Results presented here correspond to the $TIP4P/Ew$  model.}  
\includegraphics*[scale=0.45,angle=0]{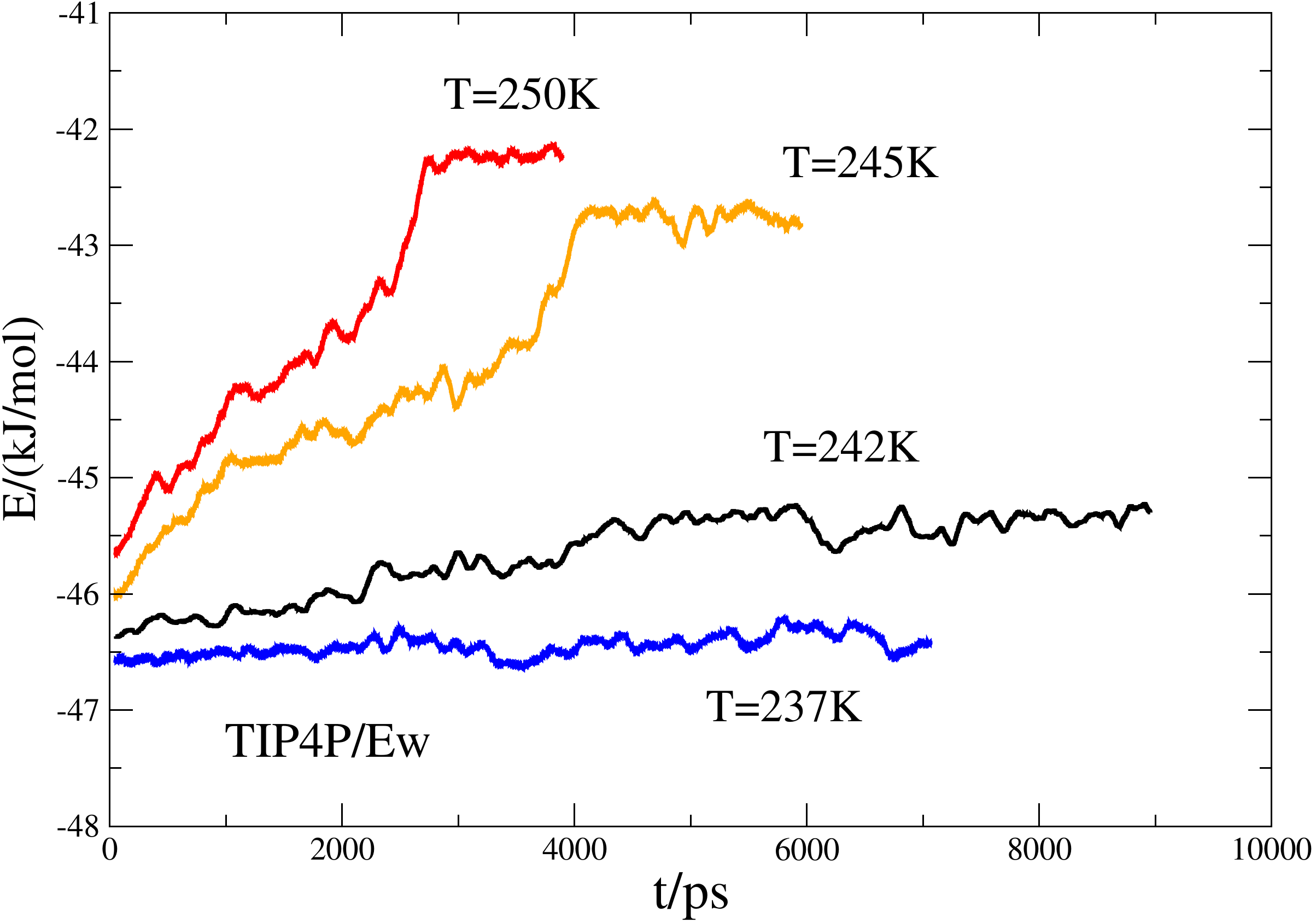} \\
\end{figure}

\newpage 
\begin{figure}[ht]
\caption{ Total energy as a function of time obtained at several 
temperatures by performing MD simulations for a block of ice $I_h$ with a free 
surface. Results presented here correspond to the $SPC/E$  model.}  
\includegraphics*[scale=0.45,angle=0]{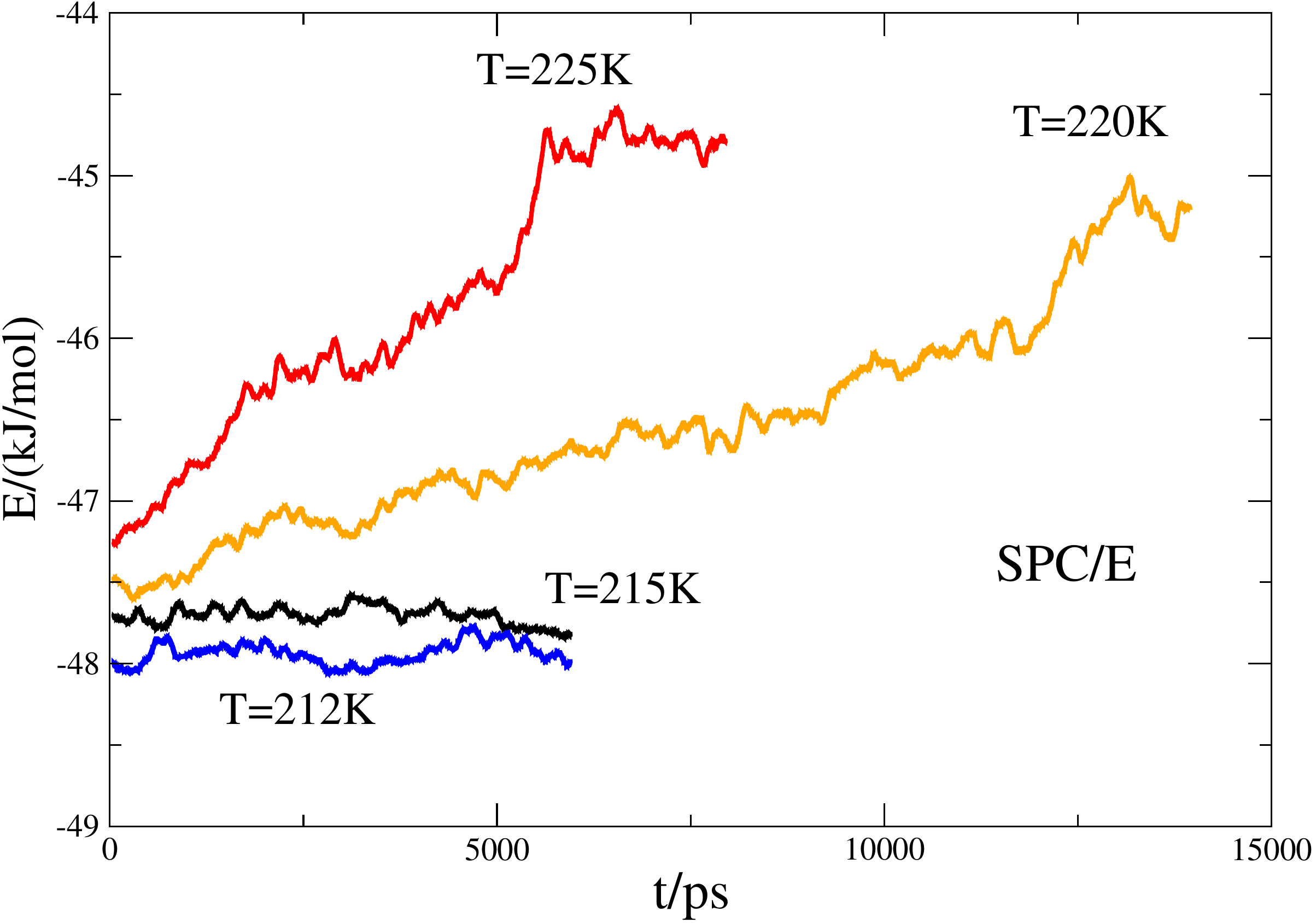} \\
\end{figure}

\newpage 
\begin{figure}[ht]
\caption{ Instantaneous configuration of the TIP4P/Ice system at 
$T=276K$ after 1ns run. }
\includegraphics[height=450pt,width=450pt,angle=0]{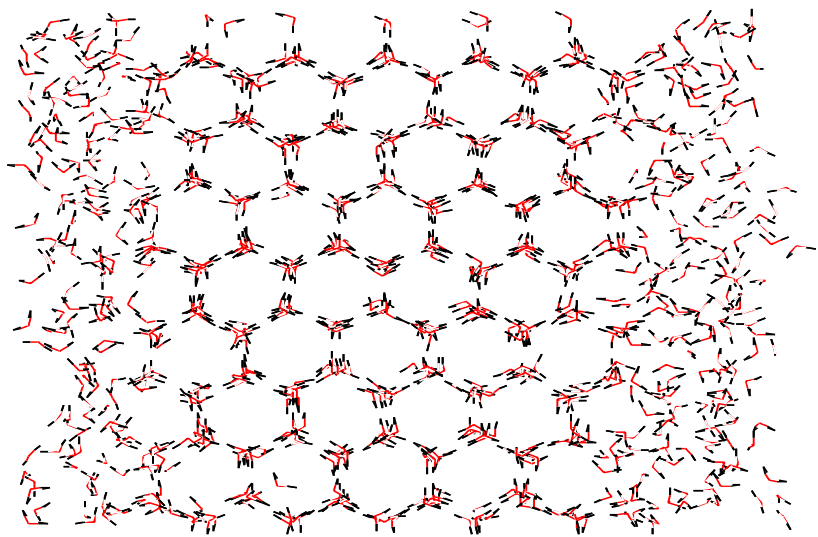}
\end{figure}

\newpage 
\begin{figure}[ht]
\caption{ Instantaneous configuration of the TIP4P/Ice system at 
$T=276K$ at the end of the run after 5ns. }
\includegraphics[height=450pt,width=450pt,angle=0]{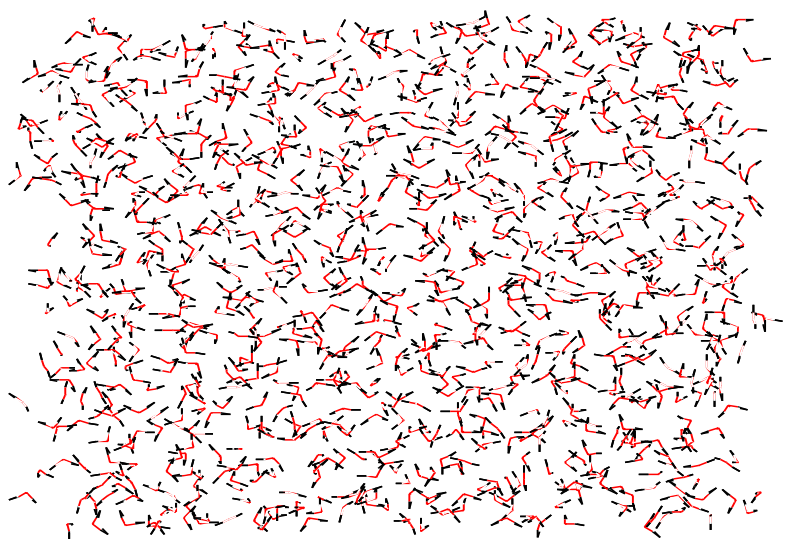}
\end{figure}

\newpage 
\begin{figure}[ht]
\caption{ Instantaneous configuration of the TIP4P/Ice system at 
$T=264K$ at the end of a 8ns run. Although the temperature is well 
below the melting point of the model, a quasi-liquid layer is clearly 
present in the ice-vacuum interface. }
\includegraphics[height=450pt,width=450pt,angle=0]{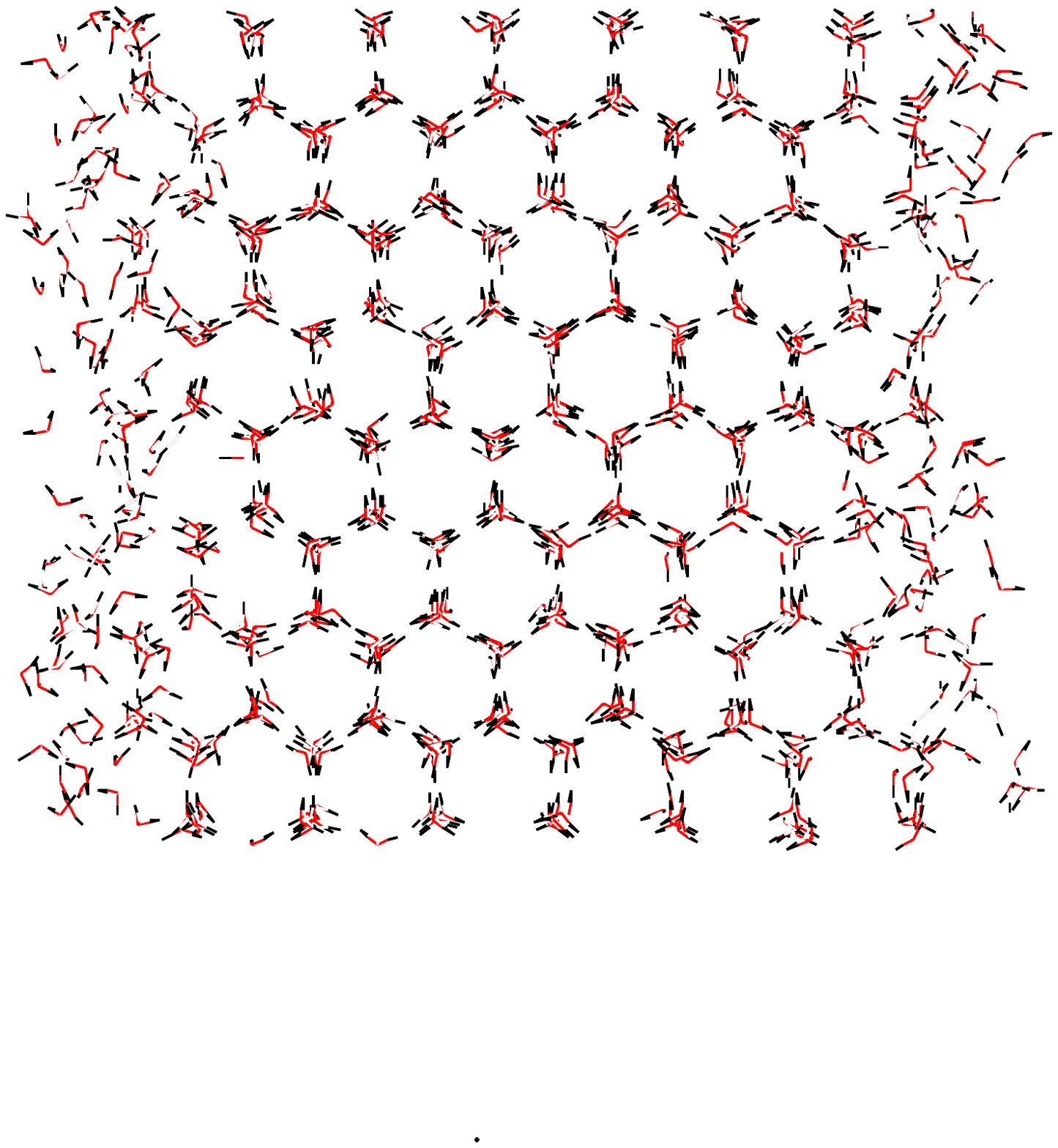}
\end{figure}

\newpage 
\begin{figure}[ht]
\caption{ Instantaneous configuration of the TIP4P/Ice system at 
$T=270K$ at the end of a 9ns run. The temperature is just below the 
melting temperature of the model. A quasi-liquid layer is clearly 
present in the ice-vacuum interface. }
\includegraphics[height=450pt,width=450pt,angle=0]{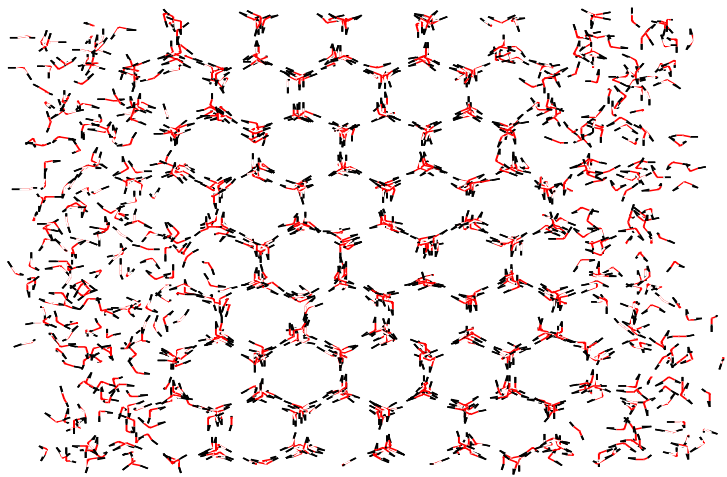}
\end{figure}
\end{document}